\documentclass[preprint,amsmath,amssymb,nofootinbib,nobibnotes, groupedaddress, prl, 11 pt]{revtex4-1}

\usepackage{graphicx}
\usepackage{dcolumn}
\usepackage{amsmath}
\usepackage{bm}
\usepackage[utf8]{inputenc}
\usepackage{mathtools}
\usepackage{pgffor}
\usepackage[normalem]{ulem}
\usepackage{xcolor}
\usepackage{BOONDOX-cal}
\usepackage{braket}
\usepackage{etaremune}

\begin{document}

    


\title{Topological phase transition between composite-fermion and Pfaffian daughter states near $\nu$ = 1/2 FQHS}
\date{\today}
    
\author{Siddharth Kumar Singh, C. Wang, C. T. Tai, C. S. Calhoun, A. Gupta, K. W. Baldwin, L. N. Pfeiffer, and M. Shayegan}
\affiliation{Department of Electrical and Computer Engineering, Princeton University, Princeton, New Jersey 08544, USA}
    
    
\begin{abstract}
\textbf{The even-denominator fractional quantum Hall state (FQHS) at filling-factor
\boldsymbol{$\nu$}=1/2 is among the most enigmatic many-body phases in two-dimensional electron systems as it appears in the ground-state rather than an excited Landau level. It is observed in wide quantum wells where the electrons have a bilayer charge distribution with finite tunneling. Whether this 1/2 FQHS is two-component (Abelian) or one-component (non-Abelian) has been debated since its experimental discovery over 30 years ago. Here, we report strong 1/2 FQHSs in ultrahigh-quality, wide, GaAs quantum wells, with transport energy gaps up to \boldsymbol{$\simeq$}4K, among the largest gaps reported for any even-denominator FQHS. The 1/2 FQHS is flanked by numerous, Jain-sequence FQHSs at \boldsymbol{$\nu$}=\boldsymbol{$p$}/(2\boldsymbol{$p$}\boldsymbol{$\pm$}1) up to \boldsymbol{$\nu$}=8/17 and 9/17. Remarkably, as we raise the density and strengthen the 1/2 FQHS, the 8/17 and 7/13 FQHSs suddenly become strong, much stronger than their neighboring high-order FQHSs. Insofar as FQHSs at \boldsymbol{$\nu$}=8/17 and 7/13 are precisely the theoretically-predicted, simplest, daughter states of the one-component Pfaffian 1/2 FQHS, our data suggest a topological phase-transition of 8/17 and 7/13 FQHSs from the Jain-states to the daughter states of the Pfaffian, and  that the parent 1/2 FQHS we observe is the Pfaffian state.}

\end{abstract}
    
\maketitle

Even-denominator FQHSs at half-filled Landau levels (LLs) are of paramount current interest as they are generally believed to possess quasi-particles that obey non-Abelian statistics and be of potential use in fault-tolerant, topological quantum computing \cite{Nayak.RMP.2008}. They are observed predominantly in two-dimensional electron systems (2DESs) at a half-filled, excited-state ($N=1$) LL, where the node in the in-plane wavefunction softens the Coulomb interaction and allows for a pairing of flux-particle composite fermions \cite{Nayak.RMP.2008, Willett.PRL.1987, Moore.Nucl.Phys.B.1991, Read.PRB.2000, Ki.Nano.Lett.2014, Li.Science.2017, Zibrov.Nature.2017, Falson.Nat.Phys.2015, Hossain.PRL.2018, Shi.NatureNano.2020, Dutta.Science.2022, Huang.PRX.2022, Willett.PRX.2023}. This is in contrast to the half-filled $N=0$ LL, namely, at LL filling factor $\nu=1/2$, where the composite fermions form a compressible Fermi sea, and the FQHSs are observed at odd-denominator fillings $\nu=p/(2p\pm1)$ on the flanks of $\nu=1/2$ (\textit{p} is an integer) \cite{Jain.Book.2007}. However, in 1992, not long after the first observation of the even-denominator FQHS in an $N=1$ LL \cite{Willett.PRL.1987}, a FQHS at $\nu=1/2$ was reported in 2DESs confined to either a single, wide, GaAs quantum well (QW) \cite{Suen.PRL.1992a}, or to a GaAs double-QW structure \cite{Eisenstein.PRL.1992}. The 1/2 FQHS in the double-QW structure, where the tunneling between the two QWs is negligible, is understood \cite{He.Das.Sarma.PRB.1993} to be the two-component, Halperin-Laughlin, $\Psi_{331}$ state \cite{Halperin.Helv.Phys.Acta.1983} which is Abelian. The origin of the 1/2 FQHS in the wide QWs, on the other hand, has been a subject of debate. Initially, it was argued to be a one-component Pfaffian state \cite{Greiter.PRB.1992, Greiter.Nucl.Phys.B.1992}, but the follow-up experiments \cite{Suen.PRL.1994, Shabani.PRB.2013}, and other theories \cite{He.Das.Sarma.PRB.1993, Peterson.PRB.2010, Thiebaut.PRB.2015} suggested a two-component state, with the components being the layer or subband degrees of freedom.  However, most recent experiments and theories favor a one-component Pfaffian state \cite{Mueed.PRL.2015, Mueed.PRL.2016, Zhu.PRB.2016, Sharma.unpublished.2023}. Evidently, the relatively large layer thickness of electrons in a wide QW softens the short-range component of the Coulomb interaction and allows for composite fermion pairing.

\begin{figure*}[t!]
\includegraphics[width=0.97\textwidth]{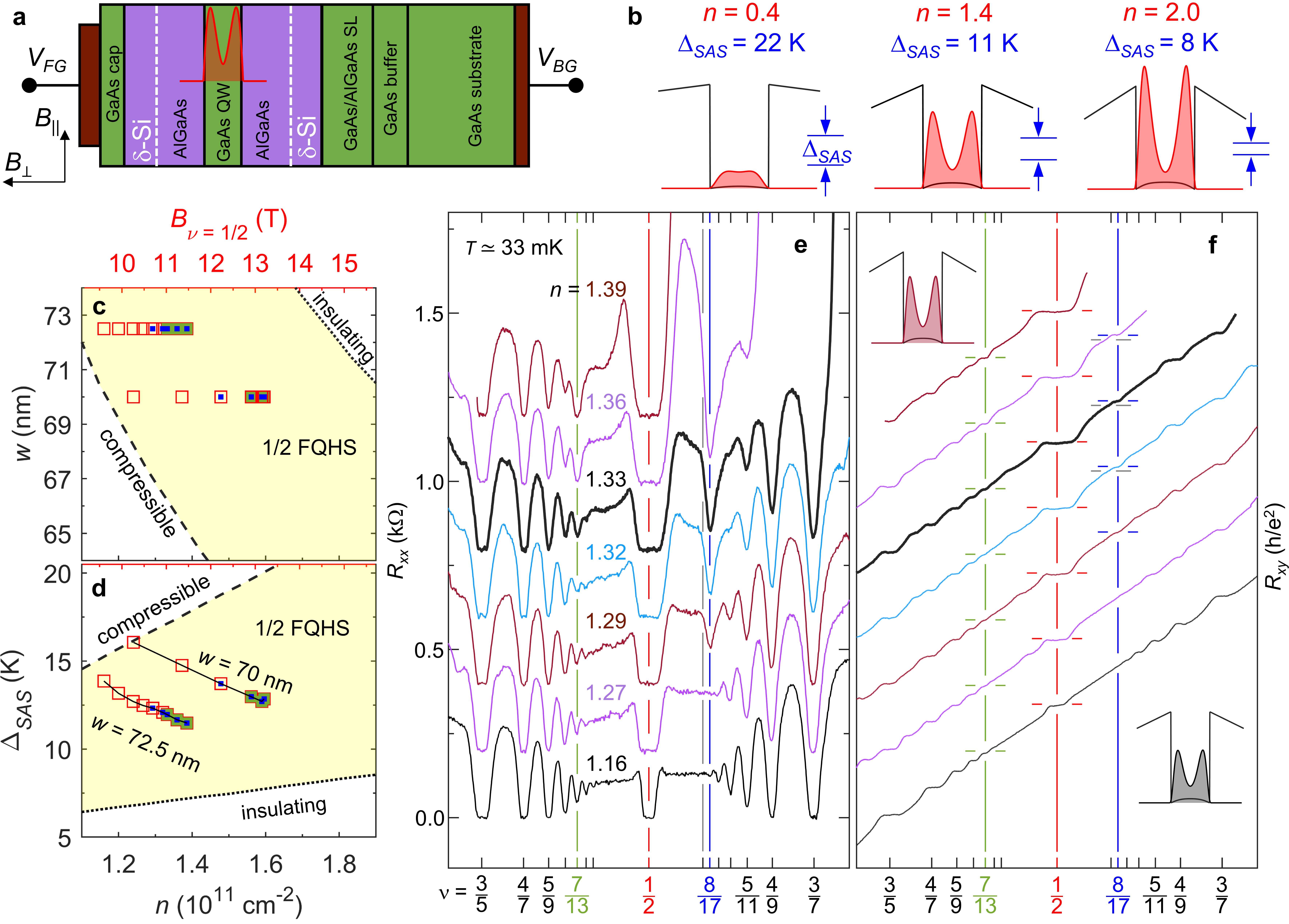}
\caption{\label{fig:fig1}\textbf{Evolution of 2D electrons in a wide QW with varying density.} \textbf{a,} Sketch of a typical, modulation-doped, GaAs QW with back and front gates which are used to tune the electron density while keeping the charge distribution symmetric. \textbf{b,} Self-consistent charge distribution (red) and potential (black) for a 2DES confined to a 72.5-nm-wide GaAs QW for different $n$ as indicated. As $n$ is increased, the 2DES becomes more “bilayer-like” with reduced interlayer tunneling ($\Delta_{SAS}$). \textbf{c, d,} Phase diagrams showing the region, marked in yellow, in which the 1/2 FQHS is stabilized for different QW widths ($w$) and $\Delta_{SAS}$ as a function of $n$. The parameters for which we observe 1/2 FQHSs in our samples agree well with these phase diagrams. Red open squares mark the points where we observe a 1/2 FQHS, while the blue and green squares indicate where we observe the anomalously strong FQHSs at $\nu=8/17$ and 7/13, respectively. \textbf{e, f,} Longitudinal ($R_{xx}$) and Hall ($R_{xy}$) resistances vs inverse filling factor of our sample at $T\simeq33$ mK for $1.16\leq n\leq 1.39$, showing the evolution of FQHSs in the $N=0$ LL. Traces are offset vertically for clarity. We normalize the x-axes by the density and show the data as a function of inverse filling factor $1/\nu=eB_\perp/nh$, where $B_\perp$ is the perpendicular magnetic field (the tick marks show the values of $\nu$). We observe anomalously strong FQHS at $\nu=8/17$ and 7/13 as $n$ is increased to 1.29 and 1.33, respectively. Green, red, blue and grey guide lines mark the expected position of $R_{xx}$ minimum and $R_{xy}$ plateau for $\nu=7/13$, 1/2, 8/17, and 9/19 respectively. The charge distribution for $n=1.16$ and 1.39 are shown as bottom and top insets in panel \textbf{f}.}
\end{figure*}

The dichotomy of the $\nu=1/2$ FQHS observed in wide GaAs QWs is highlighted in Fig. 1. Figure 1a depicts the experimental geometry, and Fig. 1b the charge distribution (red) and potential (black), calculated by self-consistently solving the Poisson and Schr\"odinger equations. The results are for electrons confined to a 72.5-nm-wide GaAs QW at three electron densities, $n=0.40$, 1.4, and 2.0, in units of $10^{11}$ cm$^{-2}$ which we use throughout this manuscript. At the smallest densities, the electrons occupy the lowest, symmetric, electric subband, and have a single-layer-like, albeit thick, charge distribution. As $n$ is increased, the inter-electron repulsion along the growth direction leads to a “bilayer-like” charge distribution \cite{Suen.PRB.1991, Suen.PRL.1992a, Suen.PRL.1994, Manoharan.PRL.1996, Shayegan.SST.Review.1996, Shabani.PRB.2013, Hatke.PRB.2017}. The electron repulsion raises the potential in the QW center and lowers the energy separation $\Delta_{SAS}$ between the symmetric and antisymmetric subbands, causing the electrons to occupy both subbands. Note that $\Delta_{SAS}$ represents the tunneling between the two electron “layers”. As shown in Fig. 1b, with increasing density, the charge distribution becomes increasingly bilayer-like and $\Delta_{SAS}$ decreases. Now, as demonstrated in Refs. \cite{Suen.PRL.1994, Manoharan.PRL.1996, Shayegan.SST.Review.1996, Shabani.PRB.2013}, the ground state at $\nu=1/2$ exhibits a remarkable evolution as the density is raised, from a compressible state to a FQHS, and then to an insulating phase. The intermediate density range where the 1/2 FQHS is stable depends critically on the QW width ($w$), leading to a $w$ vs density “phase diagram” \cite{Suen.PRL.1994, Shabani.PRB.2013}. In Fig. 1c we present an expanded section of this phase diagram, showing the FQHS region (yellow) separated from the compressible and insulating regions (white) by dashed and dotted curves, respectively. One can also highlight the stability of the 1/2 FQHS in a $\Delta_{SAS}$ vs density phase diagram, as shown in Fig. 1d. (For the full phase diagrams, see Figs. 5 and 6 of Ref. \cite{Shabani.PRB.2013}.)

\begin{figure}[t!]
\includegraphics[]{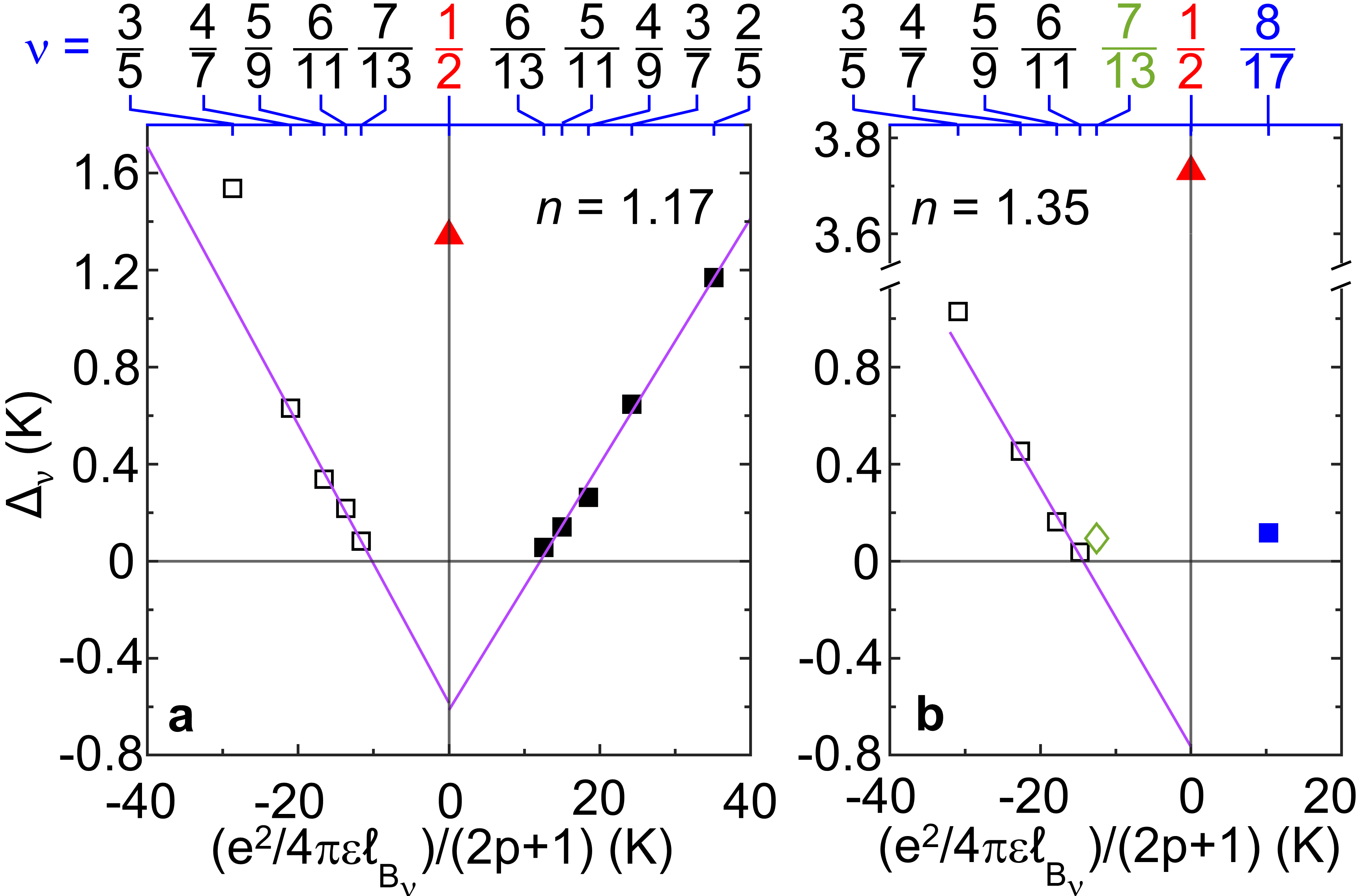}
\caption{\label{fig:fig3}\textbf{Energy gaps \boldsymbol{$\Delta_\nu$} of FQHSs in \boldsymbol{$N=0$} LL.} Measured $\Delta_\nu$ are shown as a function of $e^2/4\pi\epsilon l_{B_\nu}/(2p+1)$, where $e^2/4\pi\epsilon l_{B_\nu}$ is the Coulomb energy, $\epsilon$ is the dielectric constant of GaAs, $l_{B_\nu}$ is the magnetic length at the corresponding fillings, and $p=\nu/(1-2\nu)$. \textbf{a,} $n=1.17$, \textbf{b,} $n=1.35$. As $n$ is raised from 1.17 to 1.35, Jain-sequence FQHSs get weaker owing to the thicker 2DES except for those at $\nu=7/13$ and 8/17 which become anomalously strong. Concomitantly, the $\nu=1/2$ gap increases from 1.34 K to 3.73 K.}
\end{figure} 
    
How can these diagrams be explained? Equivalently, what is the origin of the $\nu=1/2$ FQHS observed in wide QWs? For a given $w$, at very low densities, the charge distribution is single-layer like and a compressible ground state is theoretically expected and observed at $\nu=1/2$. At very large densities, $\Delta_{SAS}$ becomes very small and the electron system essentially breaks into two layers, each with a layer filling factor 1/4. If there is no interlayer interaction, again there should be no FQHS at (total filling factor) $\nu=1/2$. If the layers are interacting and $\Delta_{SAS}$ is sufficiently small, however, the two-component, $\Psi_{331}$, FQHS can become the ground state \cite{He.Das.Sarma.PRB.1993, Peterson.PRB.2010, Thiebaut.PRB.2015}. But what about the intermediate values of $\Delta_{SAS}$? The experimental data show a very robust 1/2 FQHS even when $\Delta_{SAS}$ is about 50 K (in a GaAs QW with $w=41$ nm \cite{Shabani.PRB.2013}). It turns out, for sufficiently large $\Delta_{SAS}$, there is a fierce competitor \cite{Halperin.Surf.Sci.1994}, namely the one-component, Pfaffian FQHS \cite{Greiter.PRB.1992}. Is the 1/2 FQHS then a one-component or a two-component state? This question has been debated for over 30 years. (At very high densities and small $\Delta_{SAS}$, there is also the possibility of a bilayer Wigner crystal state; the insulating phases observed in experiments favor this possibility \cite{Manoharan.PRL.1996, Shayegan.SST.Review.1996, Hatke.PRB.2017}.)

\begin{figure*}[t!]
\includegraphics[width=1\textwidth, trim={0 0.0cm 0 0}]{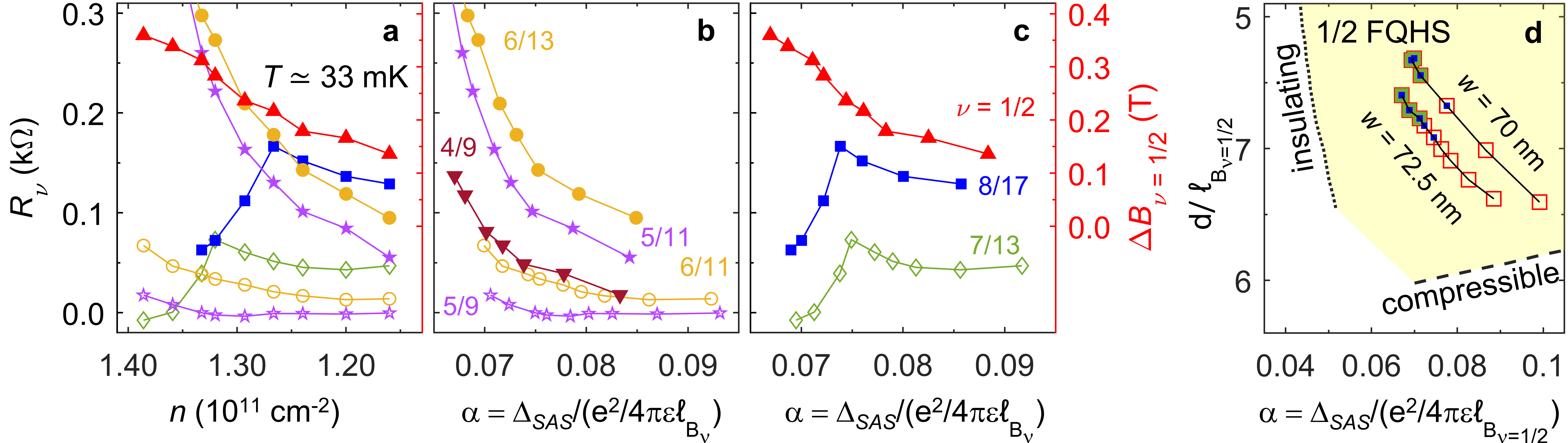}
\caption{\label{fig:fig2}\textbf{Topological phase transition in \boldsymbol{$N = 0$} LL.} \textbf{a-c,} Strengths of the hierarchical FQHSs in the $N=0$ LL, quantified by the values of their $R_{xx}$ minima are depicted. Also included in {\bf a, c} is the width of the 1/2 FQHS Hall plateau ($\Delta B_{\nu=1/2}$) as a measure of its strength. Data for different FQHSs are color coded as depicted in {\bf b, c}. In \textbf{a} data are plotted as a function of density $n$, and in \textbf{b, c,} as a function of the inter-layer tunneling  $\Delta_{SAS}$ measured in units of the Coulomb energy, $\alpha=\Delta_{SAS}/(e^2/4\pi\epsilon l_{B_\nu})$, where $l_{B_\nu}$ is the magnetic length at the corresponding fillings. We observe evidence of phase transitions from the Jain states to the 1/2 FQHS hierarchical daughter states at $\nu=8/17$ and 7/13 at $\alpha=0.075$. \textbf{d,} A $d/l_B$ vs $\alpha$ phase diagram showing where we observe the 1/2 FQHS and the anomalously strong $\nu=8/17$ and 7/13 FQHSs. Here, $d$ is the interlayer distance deduced from self-consistent calculations at zero magnetic field, and $l_B$ is the magnetic length at $\nu=1/2$. The symbols are the same as in Figs. 1c,d phase diagrams.}
\end{figure*}

The experimental data we present here shed new light on the origin of the 1/2 FQHS through the observation of its hierarchical daughter states. The traces in Figs. 1e,f show the longitudinal and Hall ($R_{xx}$ and $R_{xy}$) magnetoresistance traces for electrons confined to a 72.5-nm-wide GaAs QW at varying densities ranging from $n=1.16$ to 1.39. The lowest traces, taken at $n=1.16$, exhibit a strong FQHS, with a deep $R_{xx}$ minimum, and a well-defined and wide $R_{xy}$ plateau, quantized at $2h/e^2$. The observation of a strong 1/2 FQHS in this trace is not surprising as the parameters of the sample ($n$ and $w$) fall in the regime where a FQHS is indeed expected (see Fig. 1c phase diagram). The traces also show numerous odd-denominator FQHSs at $\nu=p/(2p\pm1)$, namely, at $\nu=1/3$, 2/5, 3/7, 4/9, $\ldots$ on the high-field side of $\nu=1/2$, and at $\nu=2/3$, 3/5, 4/7, 5/9, $\ldots$ on the low-field side. (See Extended Data Fig. 1 for trace over a bigger field range, showing FQHSs at $\nu=1/3$, 2/5, 2/3, and at other odd-denominator $\nu$.) These are the Jain-sequence FQHSs, typically seen in very high quality 2DESs with single-layer charge distributions, and can be explained as the integer QHSs of two-flux composite fermions \cite{Jain.Book.2007}. In single-layer 2DESs, the FQHSs become progressively weaker as $\nu=1/2$ is approached, and terminate in a compressible, composite fermion Fermi sea at $\nu=1/2$. The FQHS sequence observed in the $n=1.16$ trace in Fig. 1e is very similar to what is seen in single-layer systems, with the notable exception that the ground state at $\nu=1/2$ is an incompressible FQHS rather than a compressible Fermi sea. Such high-order FQHSs flanking a 1/2 FQHS have been seen previously in wide GaAs QWs but only at much higher densities \cite{Shabani.PRB.2013}. The fact that in Fig. 1e we see $R_{xx}$ minima at fractions up to $\nu=8/17$ and 9/17 ($p=8$ and 9) at a relatively low density of $n=1.16$ attests to the exceptionally high quality of the present sample \cite{Chung.NM.2021} (see Extended Data Fig. 1 and Methods for more details).
    
The striking surprise in our data comes as we increase the density in our sample. As seen in Fig. 1e, the trace at $n=1.27$ looks very similar to the $n=1.16$ trace except that the 1/2 FQHS is stronger. But at slightly higher density, $n=1.29$, the $R_{xx}$ minimum at $\nu=8/17$ suddenly becomes much deeper. In fact, it becomes deeper than the nearby, lower-order minima at $\nu=5/11$ and 6/13, and engulfs the neighboring 7/15 minimum. This behavior is in sharp contrast to the high-order FQHSs seen in the lower density traces in Fig. 1e, where the 5/11, 6/13, and 7/15 $R_{xx}$ minima are deeper than at 8/17, as expected for the standard Jain-sequence FQHSs \cite{Jain.Book.2007}. As we further increase $n$, the 8/17 $R_{xx}$ minimum becomes even deeper. At $n=1.32$ and 1.33, e.g., $R_{xx}$ minimum at 8/17 is comparable in depth to the minima at $\nu=4/9$ and 3/7. The deep 8/17 $R_{xx}$ minimum for $1.32\leq n\leq 1.36$ is accompanied by a well-developed $R_{xy}$ plateau, quantized at (17/8)($h/e^2$), as seen in Fig. 1f. It is clear that we are observing a robust FQHS at $\nu=8/17$, which is anomalously strong considering its neighboring odd-denominator states. At the highest densities in Fig. 1e, the $R_{xx}$ traces show very large values at small fillings. The 2DES starts to exhibit an insulating phase, which has been documented before and interpreted as a pinned bilayer Wigner crystal \cite{Manoharan.PRL.1996, Shayegan.SST.Review.1996, Hatke.PRB.2017}. 
    
The evolution of the FQHSs in Figs. 1e,f on the high-filling side of $\nu=1/2$ is equally impressive as we increase $n$. In this case, it is the $\nu=7/13$ FQHS that becomes anomalously strong with increasing $n$ and eventually becomes dominant over its neighboring FQHSs. It starts to strengthen as $n$ exceeds 1.32 and, at $n=1.36$ and 1.39, it exhibits a very deep $R_{xx}$ minimum and an $R_{xy}$ which is quantized at (13/7)($h/e^2$). 
    
The anomalous strengthening of the 8/17 and 7/13 FQHSs is particularly remarkable because these are exactly the simplest hierarchical daughter states of the $\nu=1/2$ Pfaffian state predicted by theory \cite{Levin.PRB.2009}. Levin and Halperin constructed a series of hierarchical daughter states for $e/4$ excitations of Pfaffian state and showed that the simplest of these states occur at $\nu=7/13$ and 8/17 \cite{Levin.PRB.2009}. In contrast, the hierarchical daughter states of the $\Psi_{331}$ state occur at $\nu=7/13$ and 9/19 \cite{Wen.AdvancePhysics.1995}, and the daughter states of the anti-Pfaffian occur at $\nu=6/13$ and 9/17 \cite{Levin.PRB.2009, Kumar.PRL.2010}. Our observation of the anomalously strong $\nu=7/13$ and 8/17 FQHSs flanking a very robust FQHS at $\nu=1/2$ provides strong evidence that the latter has a Pfaffian origin. (Note that, according to theory \cite{Levin.PRB.2009}, the 8/17 and 7/13 FQHSs are Abelian, even though their parent 1/2 FQHS is non-Abelian.)
    
It is worth noting that, in bilayer graphene where even-denominator FQHSs are observed in the $N=1$ LLs (e.g., at $\nu=3/2$), capacitance and magnetotransport studies have recently reported anomalous FQHSs on its flanks at partial filling factors $\tilde{\nu}=7/13$ and 8/17 \cite{Zibrov.Nature.2017, Li.Science.2017, Huang.PRX.2022}, and have been interpreted as the hierarchical daughter states of the even-denominator Pfaffian FQHS. A crucial distinction is that, in our wide GaAs QW sample, we observe the $\nu=1/2$ FQHS and its hierarchical daughter FQHSs at 7/13 and 8/17 in the \textit{ground} ($N=0$) LL rather than in the \textit{excited} ($N=1$) LL. Moreover, in bilayer graphene, no transition is reported at these fillings. 
    
Next, we focus on the evolution of the FQHSs seen in Figs. 1e,f more closely. We begin again with the lowest density, $n=1.16$ trace. The Jain-sequence FQHSs observed at $\nu=p/(2p\pm1)$ in this trace have the same pattern as in the standard, single-layer 2DESs, namely, they monotonically become weaker as $p$ increases. This trend can be seen quantitatively in Fig. 2a where we plot the energy gaps $\Delta_\nu$, determined from the temperature dependence of the resistance minimum $R_{xx}\propto e^{-\mathbin{\Delta_\nu}/2kT}$; see Extended Data Fig. 2. For an ideal 2DES (with zero layer thickness, no LL mixing, and no disorder), these gaps are expected to scale as: $\Delta_{\nu}=(C/|2p+1|)E_{C}$ \cite{HLR.PRB.1993,Jain.Book.2007}, where $E_C=e^2/4\pi\epsilon l_{B_\nu}$ is the Coulomb energy, $l_{B_{\nu}}=\sqrt{\hbar/\textit{e}B_{\nu}}$ is the magnetic length, $C\simeq0.3$, and $\nu=p/(2p+1)$. As seen in Fig. 2a, the gaps monotonically get smaller for higher $p$ until they become immeasurably small. This trend is similar to what is experimentally observed and theoretically expected in narrower GaAs QWs \cite{Du.PRL.1993, Manoharan.PRL.1994, Villegas.Rosales.PRL.2021, Zhao.PRB.2022}. Note that the measured gaps near $\nu=1/2$ extrapolate to a negative value at $\nu=1/2$. This is consistent with previous measurements where the negative intercept is interpreted to reflect broadening due to the sample disorder \cite{Du.PRL.1993, Manoharan.PRL.1994, Villegas.Rosales.PRL.2021}. Also consistent with previous studies are the smaller values of the measured gaps in Fig. 2a compared to narrower QWs; the larger effective thickness of the electron layer in the present sample softens the short-range electron-electron interaction and leads to smaller FQHS gaps \cite{Villegas.Rosales.PRL.2021,Zhao.PRB.2022}. What is of course unusual in Fig. 2a and the $R_{xx}$ trace in Fig. 1e is that there is a FQHS at $\nu=1/2$, even though the high-order FQHSs on its flanks behave the same as in a standard, albeit thick, 2DES.
    
As the density is raised, the traces in Fig. 1e initially reveal a gradual weakening of all the $\nu=p/(2p\pm1)$ FQHSs. This is best seen in Fig. 3a which shows that values of $R_{xx}$ minima gradually increase as {\it n} is raised, and can be explained to result from the increase in the electron system's effective thickness \cite{Villegas.Rosales.PRL.2021}. As the density is further raised, the $\nu=p/(2p\pm1)$ FQHSs continue to become weaker as evinced from the increase in their $R_{xx}$ minima values (Fig. 3a), except of course the FQHSs at $\nu=8/17$ and 7/13, which suddenly start to strengthen at $n=1.29$ and 1.33, respectively. (The 8/17 FQHS weakens and disappears at the highest $n$ as an insulating phase engulfs it.) If we convert the x-axis in Fig. 3a to $\alpha=\Delta_{SAS}/(e^2/4\pi\epsilon l_{B_\nu})$, i.e., a dimensionless parameter quantifying the ratio of the tunneling energy to the in-plane interaction energy, as is commonly done in numerous studies \cite{Suen.PRL.1994,Peterson.PRB.2010, Thiebaut.PRB.2015, Zhu.PRB.2016, Manoharan.PRL.1996, Shayegan.SST.Review.1996, Halperin.Surf.Sci.1994}, we find that the sudden strengthening of the 8/17 and 7/13 FQHSs indeed occurs essentially at the same value of $\alpha=0.075$ (Figs. 3b,c). 
    
The data of Figs. 3b,c reveal a clear, sudden transition at 8/17 and 7/13 from Jain-sequence FQHSs to FQHSs that are predicted to be the daughter states of the Pfaffian state \cite{Levin.PRB.2009}. The theory of Levin and Halperin \cite{Levin.PRB.2009} also posits that the two sets are topologically distinct. Our data therefore suggest that we are seeing a topological phase transition at $\nu=7/13$ and 8/17 from Jain-sequence FQHSs to the hierarchical daughter FQHSs of the Pfaffian 1/2 FQHS in our sample. 

It is remarkable that in the whole density range of our experiments on this sample, even as the 8/17 and 7/13 FQHSs are undergoing a phase transition, the 1/2 FQHS monotonically becomes stronger as $n$ increases. This is evidenced by the steady increase of the $R_{xy}$ plateau width ($\Delta B_{\nu}$) with density (Figs. 3a,c). It is possible that the stability of the 8/17 and 7/13 FQHSs as the daughter states of the 1/2 FQHS requires a minimal strength of the latter. As Fig. 2b illustrates, the 1/2 FQHS at $n=1.35$ is indeed extremely robust. It has an energy gap of $\simeq 3.73$ K; this is much larger than the transport gaps reported for any even-denominator FQHS in any semiconducting material platform \cite{Chung.NM.2021, Falson.Nat.Phys.2015, Shi.NatureNano.2020}, and comparable to the largest gaps reported in bilayer graphene \cite{Li.Science.2017, Zibrov.Nature.2017, Huang.PRX.2022}. This is particularly remarkable, considering that the size of our sample ($\simeq 16$ mm$^{2}$) is about 10$^{6}$-10$^{7}$ larger than the typical bilayer graphene samples; the much larger size is important for potential use of these materials as platforms for topological quantum computing. 

\begin{figure}[t!]
\includegraphics[]{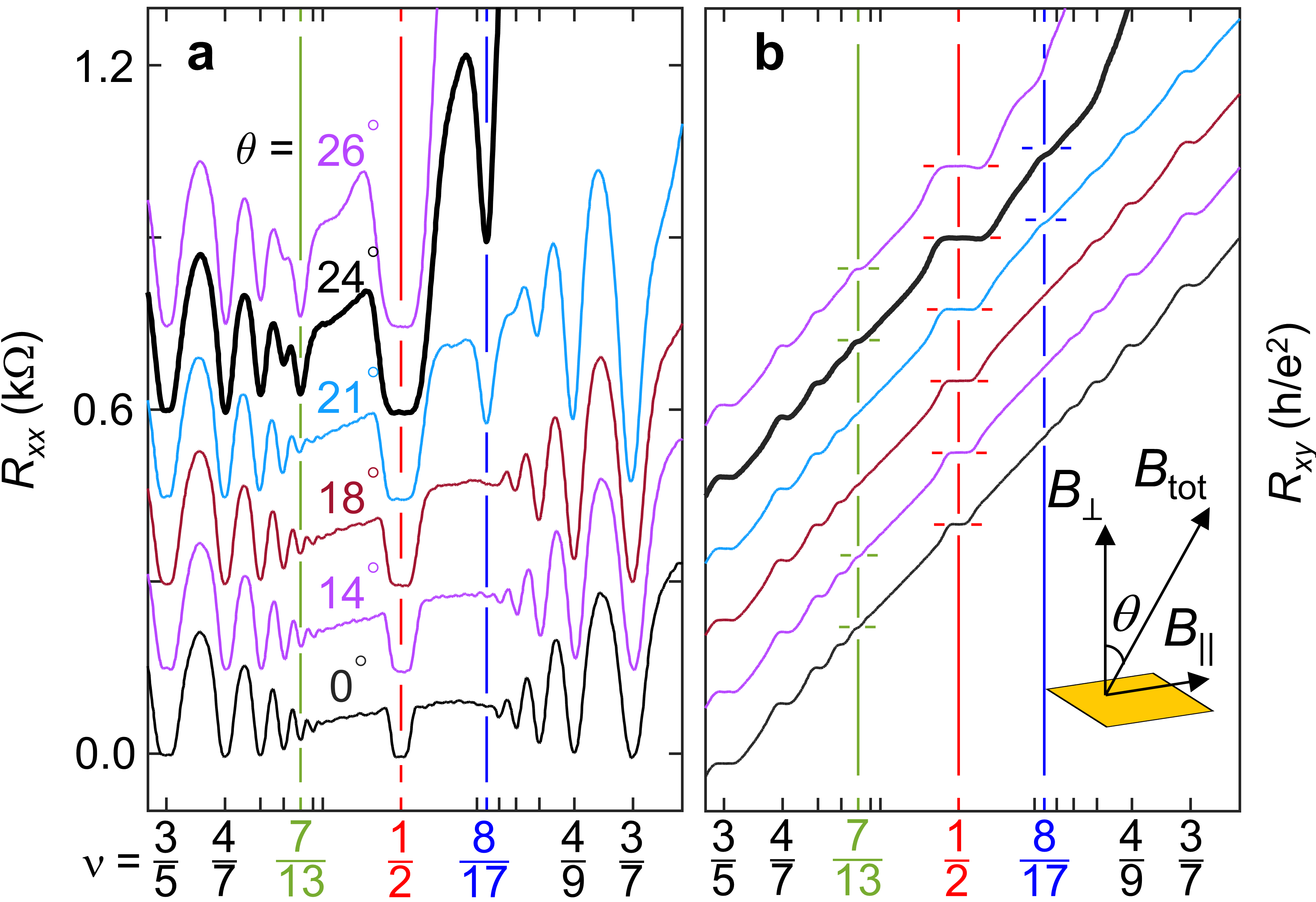}
\caption{\label{fig:fig4}\textbf{Effect of in-plane magnetic field}. \textbf{a, b,} $R_{xx}$ and $R_{xy}$ show qualitatively similar behavior to increasing the 2DES density. Traces have been offset vertically for clarity. Inset in {\bf b} shows the angle $\theta$.}
\end{figure}
    
Finally, we present the evolution of the FQHSs in our sample by tilting it in the magnetic field and introducing a parallel field component ($B_{||}$); see Fig. 4b inset. Thanks to its coupling to the out-of-plane orbital motion of electrons, $B_{||}$ reduces the interlayer tunneling, qualitatively similar to increasing $n$ \cite{Lay.PRB.1997, Hasdemir.PRB.2015}. Figure 4 reveals that when we tilt the sample at a low density where the strengths of FQHSs follow the Jain-sequence protocol at $\theta=0$, anomalously strong FQHSs appear at 8/17 and 7/13 at higher $\theta$. The fact that the 8/17 FQHS shows up at a lower $\theta$ compared to the 7/13 FQHS is qualitatively consistent with the density-dependence data of Figs. 1e,f; at a given $\theta$, the tunneling is more suppressed at the field position of the 8/17 FQHS compared to the 7/13 because of the larger $B_{||}$. It is worth mentioning that, although not highlighted, the tilt data of Ref. \cite{Hasdemir.PRB.2015} also exhibit clear hints of anomalously-strong 8/17 and 7/13 FQHSs (see traces at $\theta=35^\circ$ and $37^\circ$ in Fig. 1 and $10^\circ$ and $20^\circ$ in Fig. 2(a) of \cite{Hasdemir.PRB.2015}).
    
Our data presented here provide evidence that the $\nu=8/17$ and 7/13 FQHSs flanking the $\nu=1/2$ FQHS in a 72.5-nm-wide GaAs QW make transitions from the Jain states to anomalously strong FQHSs as the density is increased and the $\nu=1/2$ FQHS is made stronger. The abruptness of these these transitions (see e.g. Fig. 3c) suggest that they are first order. We also made measurements on a narrower (70-nm-wide) GaAs QW; see Extended Data Figs. 3-5. The results confirm our findings for the wider QW and show similar phase transitions, albeit at slightly larger densities, as summarized in the phase diagrams shown in Figs. 1c,d and 3d. There is little question that the exceptional quality and purity of the samples is crucial for our observations (see Methods). However, the precise parameters (QW width, density, $\Delta_{SAS}$, etc.) contained in these phase diagrams are also of paramount importance as they provide unique and precise experimental input for future studies, both experimental and theoretical, to further unravel the physics of the FQHSs in wide QWs. 

\textit{Note added} -- After the completion of our work, we became aware of two, recent preprints reporting capacitance, transport, and tunneling data near the even-denominator FQHSs in bilayer graphene \cite{Assouline.Preprint.2023, Hu.Preprint.2023}. They report large energy gaps for the even-denominator FQHSs, and confirm the observation of hierarchical daughter states at partial LL filling factors $\tilde{\nu}=8/17$ and 7/13 in the $N=1$ LL in bilayer graphene \cite{Zibrov.Nature.2017, Li.Science.2017, Huang.PRX.2022}. Also, in Fig. 4 of Ref. \cite{Hu.Preprint.2023}, data are shown near the half-filled $N=2$ LL. The data have a striking resemblance to the data seen in the $N=0$ LL in GaAs wide QW samples at low densities before the 8/17 and 7/13 become anomalously strong, namely, they exhibit a relatively robust even-denominator FQHS flanked by numerous odd-denominator FQHSs at $\tilde{\nu}=p/(2p\pm1)$ whose strengths follow the Jain-sequence protocol (see, e.g., the traces at $n=1.16$ in Figs. 1e, f and Fig. 5 in Appendix B). This behavior for the $N=0$ LL in GaAs wide QWs has of course been well established and documented in numerous previous reports \cite{Suen.PRL.1994, Shabani.PRB.2013}. What is surprising is that it is now seen in the $N=2$ LL of bilayer graphene.

\section{Methods}
    
Our samples, grown using molecular beam epitaxy (MBE), exhibit record-high mobilities of $\mu\simeq10\times10^{6}$ cm\textsuperscript{2}V\textsuperscript{-1}s\textsuperscript{-1} for 2DESs confined to wide GaAs QWs where two subbands are occupied. This record mobility was achieved by improvements in MBE vacuum integrity, extensive Al and Ga source purification and doping the sample with Si in doping-well structures \cite{Chung.NM.2021}.

As discussed in the main text, and also in Refs. \cite{Suen.PRB.1991, Suen.PRL.1992a, Suen.PRL.1994, Manoharan.PRL.1996, Shayegan.SST.Review.1996, Shabani.PRB.2013}, the 2DES in a wide GaAs QW is a particularly flexible and versatile platform as its important parameters, such as the shape of the charge distribution and inter-layer tunneling, can be tuned over a wide range (Fig. 1b). As demonstrated in this study, changes in charge distribution profoundly affect the correlated states of the wide QW, and particularly the $\nu=1/2$ FQHS and its daughter states. It is worth highlighting three main advantages of electrons confined to a wide QW over those in a double-QW structure, e.g., electrons confined to two GaAs QWs separated by an Al$_{x}$Ga$_{1-x}$As barrier. First, the potential controlling the tunneling and the (bilayer) charge distribution in wide QWs is generated by the electrons themselves and can be tuned over a significant range by changing the electron density. Second, in a double-QW structure, the interfaces between the GaAs QWs and the Al$_{x}$Ga$_{1-x}$As barriers can lead to interface roughness scattering. Third, when inter-layer tunneling is substantial in a double-QW structure, the electron charge distribution penetrates deep into the Al$_{x}$Ga$_{1-x}$As barrier, causing alloy scattering. Moreover, the purity of the Al$_{x}$Ga$_{1-x}$As barrier is not as high as that a pure GaAs layer \cite{Chung.NM.2021}. As a result, the electrons in a wide QW suffer much less from disorder, as indeed confirmed by their exceptionally high mobilities, and the display of delicate, exotic, many-body states.

Samples are prepared for magnetotransport experiments by cleaving a 4 mm $\times$ 4 mm square piece from a 2-inch GaAs wafer which has the MBE-grown structure. Eutectic In:Sn is deposited on the sample with a soldering iron and subsequently annealed at 425 $^\circ$C for 4 minutes in a reducing environment of 95$\%$:5$\%$ N\textsubscript{2}:H\textsubscript{2} for making reliable Ohmic contacts at sample's four corners and side midpoints. Front-gate consists of a 10-nm-thick layer of Ti plus a 30-nm-thick layer of Au which are deposited under high vacuum using an electron-beam evaporator. Indium is melted on the back-surface to serve as the back-gate electrode. 

Measurements are performed in van der Pauw geometry using standard lock-in techniques. All measurements at base temperature $T\simeq33$ mK were made using 20 nA excitation current; the energy gap measurements were made using 50 nA excitation current.
    
\section{Data availability}
    
The data that support the plots within this paper and other findings of this study are available from the corresponding author upon reasonable request.
    
\section{Acknowledgments}
    
We acknowledge support by the U.S. Department of Energy Basic Energy (DOE) Sciences Grant No. DEFG02-00-ER45841) for measurements, the National Science Foundation (NSF) Grants Nos. DMR 2104771 and ECCS 1906253) for sample characterization, and the Eric and Wendy Schmidt Transformative Technology Fund and the Gordon and Betty Moore Foundation’s EPiQS Initiative (Grant No. GBMF9615 to L.N.P.) for sample fabrication. Our measurements were partly performed at the National High Magnetic Field Laboratory (NHMFL), which is supported by the NSF Cooperative Agreement No. DMR 1644779, by the State of Florida, and by the DOE. This research is funded in part by QuantEmX grant from Institute for Complex Adaptive Matter and the Gordon and Betty Moore Foundation through Grant GBMF9616 to S.K.S., C. W., A. G., C.T.T. and C.S.C. We thank R. Nowell, G. Jones, Ali Bangura and T. Murphy at NHMFL for technical assistance, and Bert I. Halperin, Jainendra K. Jain, Prashant Kumar, Michael Levin, and Xiao-Gang Wen for illuminating discussions.
    
\section{Author contributions}
    
S.K.S. and M.S. conceived the work. S.K.S., C.W., C.T.T. and C.C. performed low temperature transport measurements. S.K.S. and K.W.B. fabricated the sample. L.N.P., K.W.B. and A.G. produced molecular beam epitaxy samples and characterized them. S.K.S. and M.S. analysed the data and wrote the manuscript with input from all co-authors. 
    
\section{Competing interests}
    
The authors declare no competing interests.
    
\section{Extended data}
    
\setcounter{figure}{0}
\renewcommand{\figurename}{Extended Data Fig.}
    
\begin{figure*}[h!]
\includegraphics[width=1\textwidth, trim={0 0.0cm 0 0}]{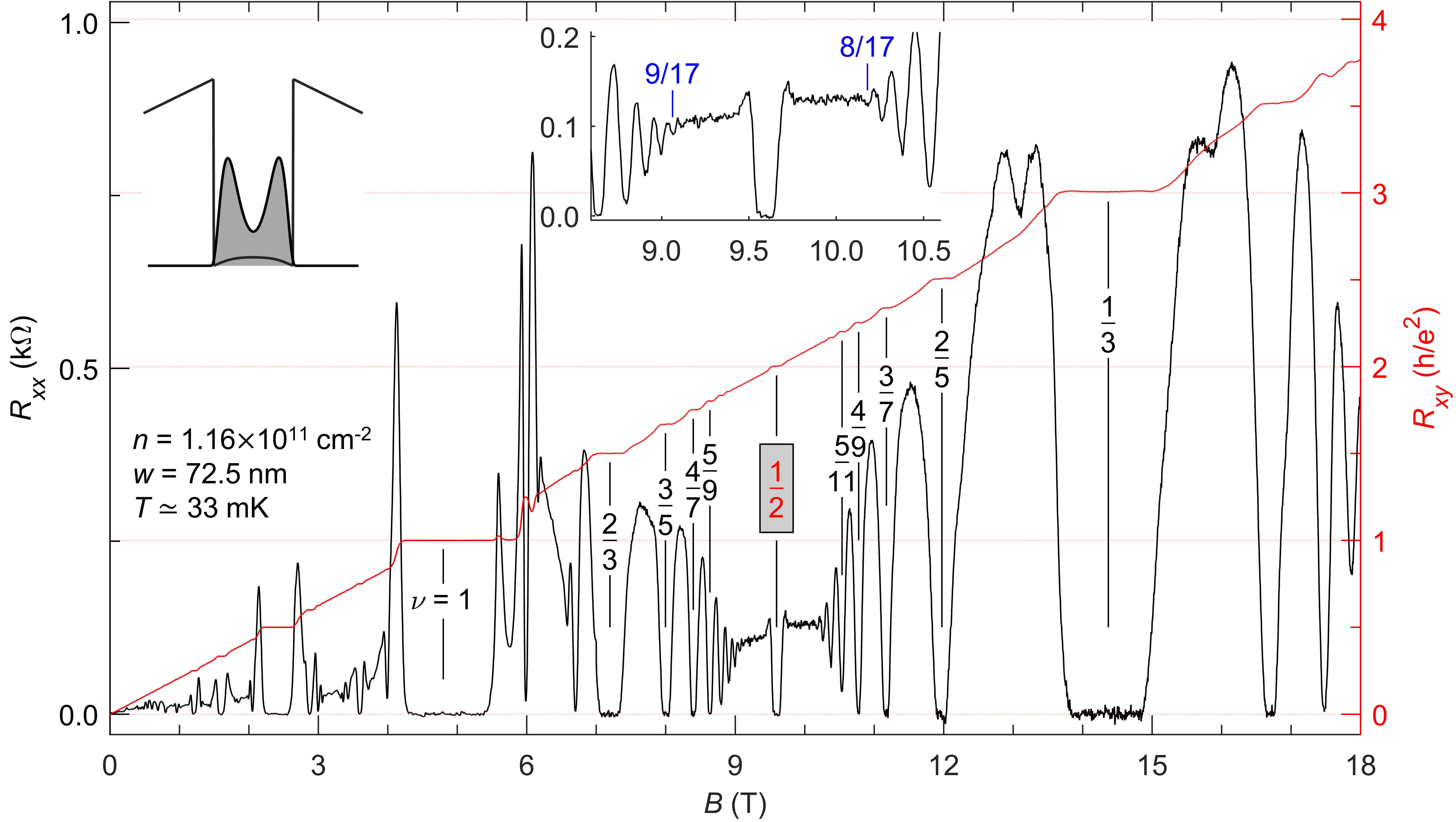}
\caption{\label{fig:e1}\textbf{Longitudinal (\boldsymbol{$R_{xx}$}) and Hall resistance (\boldsymbol{$R_{xy}$}) as a function of perpendicular magnetic field for ultrahigh quality 2D electrons in a 72.5-nm-wide GaAs QW.} The strong FQHS observed at $\nu=1/2$ flanked by numerous high-order Jain-sequence FQHSs up to $\nu=8/17$ and 9/17 (as shown in the top right inset) on its sides attest to the ultra high quality of our sample. Top left inset shows the self-consistently calculated charge distribution for $n=1.16\times10^{11}$ cm$^{-2}$ electrons confined to a 72.5-nm-wide GaAs QW.} 
\end{figure*}
    
\begin{figure*}[t!]
\includegraphics[width=1\textwidth, trim={0 0 0 1.1in},clip]{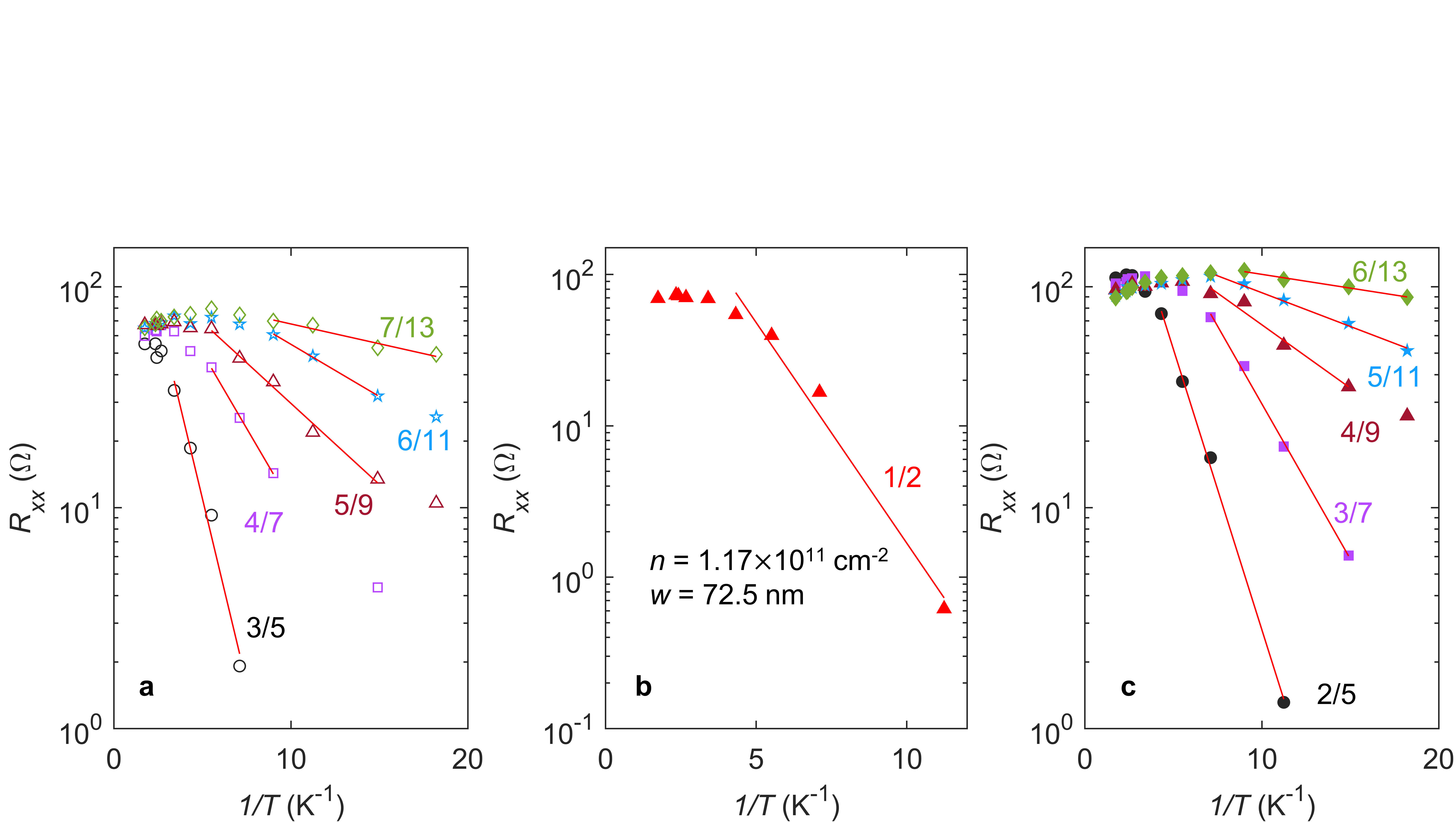}
\caption{\label{fig:e2}\textbf{Arrhenius plots to extract the energy gaps of FQHSs in the \boldsymbol{$N=0$} LL.} Data are for the 72.5-nm-wide GaAs QW with $n=1.17\times10^{11}$ cm$^{-2}$. {\bf a-c,} Temperature dependence of $R_{xx}$ minima for the FQHSs on the low-field side of $\nu=1/2$, at $\nu=1/2$, and on the high-field side of $\nu=1/2$, respectively. The red lines through the data points are fits to the data points in the activated regimes for different fillings, and their slopes yield the energy gaps $\Delta_\nu$, determined from $R_{xx}\propto e^{-\mathbin{\Delta_\nu}/2kT}$. Note in panel {\bf a} that the $\nu=7/13$ FQHS is weaker than the 6/11 FQHS and has a smaller energy gap.}
\end{figure*}
    
\begin{figure*}[t!]
\includegraphics[width=1\textwidth, trim={0 0 0 1.1in},clip]{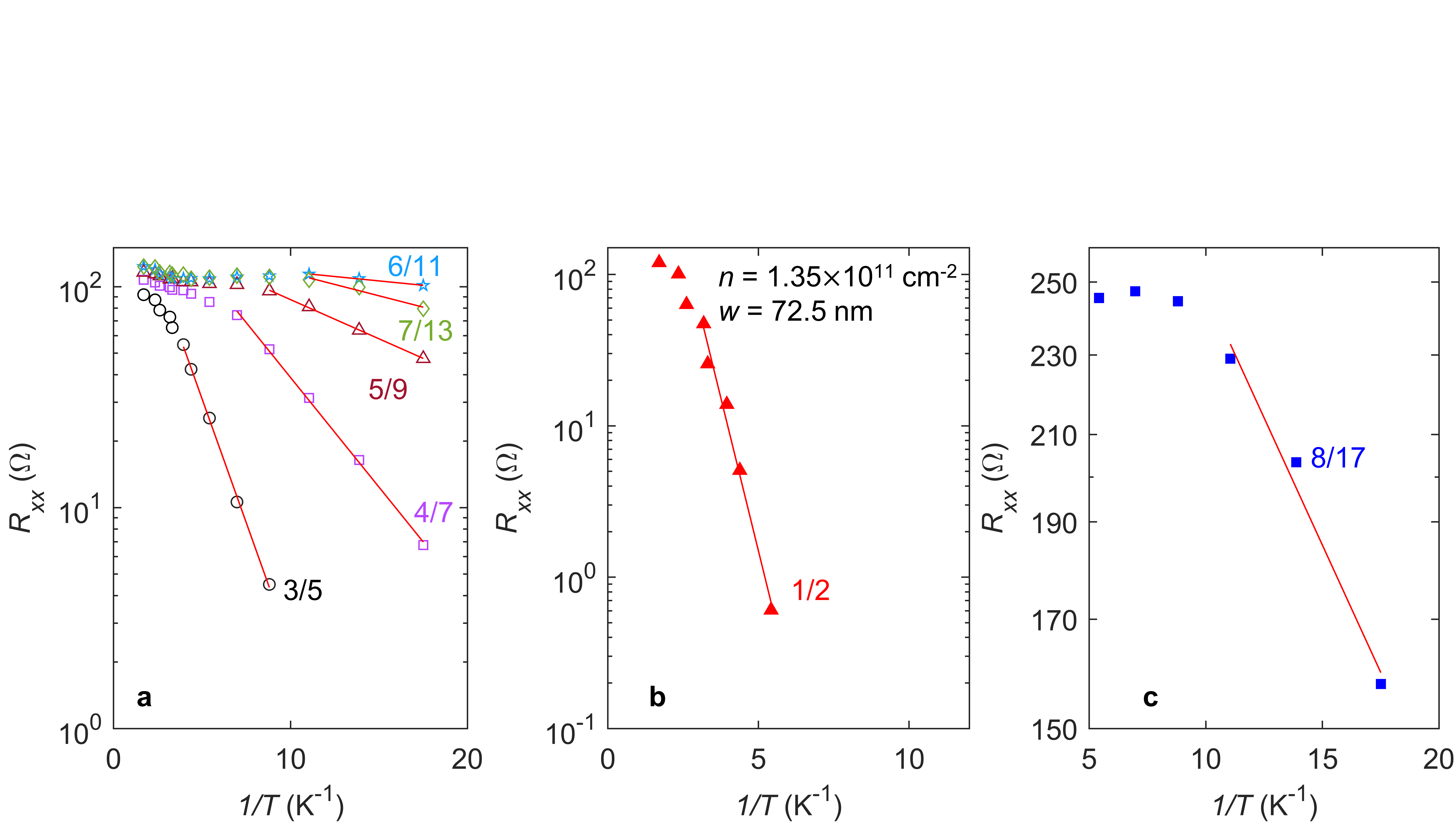}
\caption{\label{fig:e3}\textbf{ Arrhenius plots to extract the energy gaps of FQHSs in the \boldsymbol{$N=0$} LL.} Data are for the 72.5-nm-wide GaAs QW with $n=1.35\times10^{11}$ cm$^{-2}$. {\bf a-c,} Temperature dependence of $R_{xx}$ minima for the FQHSs on the low-field side of $\nu=1/2$, at $\nu=1/2$, and on the high-field side of $\nu=1/2$, respectively. The red lines through the data points are fits to the data points in the activated regimes for different fillings, and their slopes yield the energy gaps $\Delta_\nu$, determined from $R_{xx}\propto e^{-\mathbin{\Delta_\nu}/2kT}$. Note in panel {\bf a} that, at this density, the 7/13 FQHS has a larger gap than the 6/11 FQHS. On the high-field side of $\nu=1/2$ (panel {\bf c}), only the $\nu=8/17$ FQHS yields an activation, and the other FQHSs at higher fields are consumed by the ensuing insulating phase; see Fig. 1e of main text.}
\end{figure*}
    
\begin{figure*}[t!]
\includegraphics[width=1\textwidth, trim={0 0.0cm 0 0}]{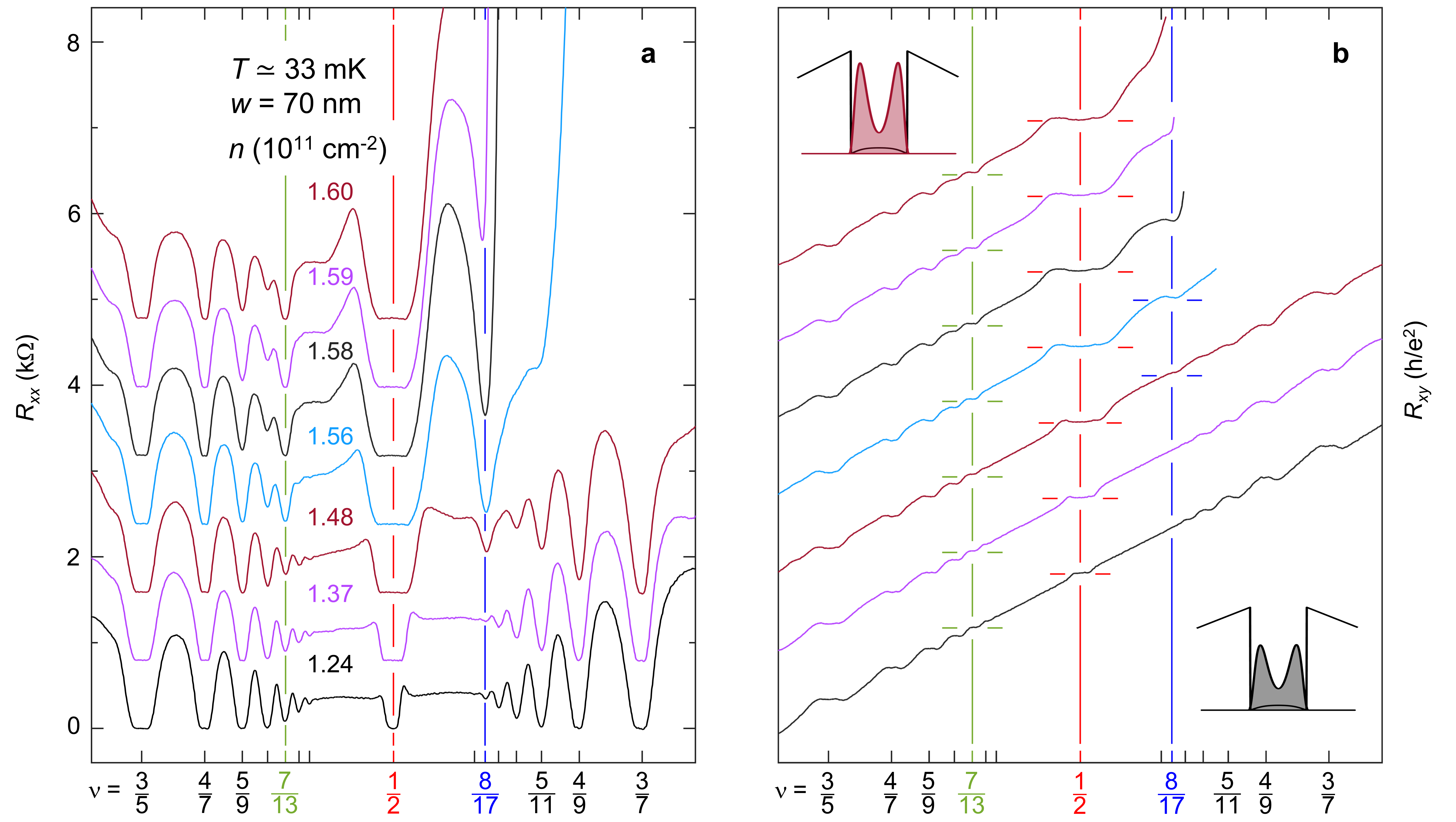}
\caption{\label{fig:e4}\textbf{Evolution of an electron system in a 70-nm-wide GaAs QW with varying density.} {\bf a,b,} Longitudinal ($R_{xx}$) and Hall resistance ($R_{xy}$) as a function of $1/\nu$ in a 70 nm wide GaAs QW. We see a similar evolution of FQHSs at and around $\nu=1/2$ as in the 72.5-nm sample. Quantitavely, owing to the smaller well width, the transitions of the $\nu=8/17$ and 7/13 FQHSs are both shifted to higher densities compared to the 72.5-nm-wide sample of Fig. 1. In panel {\bf b}, self-consistently calculated charge distribution for a 2DES confined to a 70-nm-wide GaAs QW at $n=1.24$ and $1.60\times10^{11}$ cm$^{-2}$ are shown as bottom and top insets, respectively.}
\end{figure*}
    
\begin{figure*}[t!]
\includegraphics[width=1\textwidth, trim={0 0.0cm 0 0}]{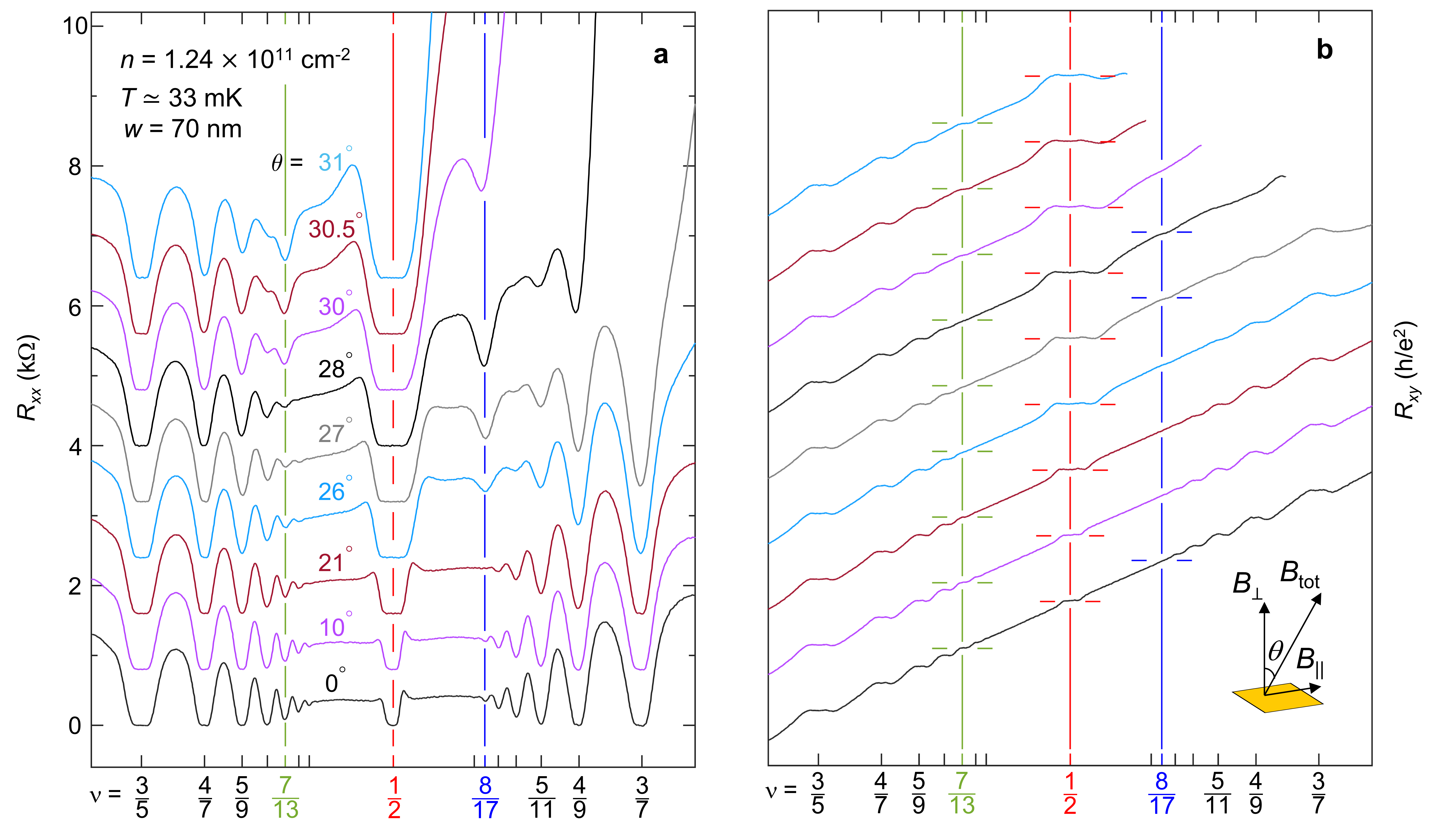}
\caption{\label{fig:e5}\textbf{Evolution of an electron system in a 70-nm-wide GaAs QW with in-plane magnetic field.} {\bf a,b,} Longitudinal ($R_{xx}$) and Hall ($R_{xy}$) resistances are shown as a function of $1/\nu$ at various tilt angles for $n=1.24\times10^{11}$ cm$^{-2}$. Qualitatively, once again the sample shows a similar evolution of FQHSs at and around $\nu=1/2$ as the 72.5-nm-wide sample, albeit at higher tilt angles.}
\end{figure*}
    
\end{document}